\documentstyle[12pt]{article}
\def\Re{{\cal R \mskip-4mu \lower.1ex \hbox{\it e}\,}}
\def\Im{{\cal I \mskip-5mu \lower.1ex \hbox{\it m}\,}}
\def\ie{{\it i.e.}}
\def\eg{{\it e.g.}}

\def\etal{{\it et al.}}
\def\ibid{{\it ibid}.}
\def\sub#1{_{\lower.25ex\hbox{$\scriptstyle#1$}}}
\def\sul#1{_{\kern-.1em#1}}
\def\sll#1{_{\kern-.2em#1}}
\def\sbl#1{_{\kern-.1em\lower.25ex\hbox{$\scriptstyle#1$}}}
\def\ssb#1{_{\lower.25ex\hbox{$\scriptscriptstyle#1$}}}
\def\sbb#1{_{\lower.4ex\hbox{$\scriptstyle#1$}}}

\def\to{\rightarrow}
\def\mh{\ifmmode m\sbl H \else $m\sbl H$\fi}
\def\mch{\ifmmode m_{H^\pm} \else $m_{H^\pm}$\fi}
\def\mt{\ifmmode m_t\else $m_t$\fi}
\def\mc{\ifmmode m_c\else $m_c$\fi}
\def\mz{\ifmmode M_Z\else $M_Z$\fi}
\def\mw{\ifmmode M_W\else $M_W$\fi}
\def\mws{\ifmmode M_W^2 \else $M_W^2$\fi}
\def\mhs{\ifmmode m_H^2 \else $m_H^2$\fi}
\def\mzs{\ifmmode M_Z^2 \else $M_Z^2$\fi}
\def\mts{\ifmmode m_t^2 \else $m_t^2$\fi}
\def\mcs{\ifmmode m_c^2 \else $m_c^2$\fi}
\def\mchs{\ifmmode m_{H^\pm}^2 \else $m_{H^\pm}^2$\fi}
\def\ztwo{\ifmmode Z_2\else $Z_2$\fi}
\def\zone{\ifmmode Z_1\else $Z_1$\fi}
\def\mtwo{\ifmmode M_2\else $M_2$\fi}
\def\mone{\ifmmode M_1\else $M_1$\fi}
\def\tb{\ifmmode \tan\beta \else $\tan\beta$\fi}
\def\xw{\ifmmode x\sub w\else $x\sub w$\fi}
\def\ch{\ifmmode H^\pm \else $H^\pm$\fi}
\def\lum{\ifmmode {\cal L}\else ${\cal L}$\fi}
\def\inpb{\ifmmode {\rm pb}^{-1}\else ${\rm pb}^{-1}$\fi}
\def\infb{\ifmmode {\rm fb}^{-1}\else ${\rm fb}^{-1}$\fi}
\def\epem{\ifmmode e^+e^-\else $e^+e^-$\fi}
\def\ppb{\ifmmode \bar pp\else $\bar pp$\fi}

\def\bsg{\ifmmode b\rightarrow s\gamma \else $b\rightarrow s\gamma$\fi}

\newskip\zatskip \zatskip=0pt plus0pt minus0pt
\def\matth{\mathsurround=0pt}

\def\atversim#1#2{\lower0.7ex\vbox{\baselineskip\zatskip\lineskip\zatskip
  \lineskiplimit 0pt\ialign{$\matth#1\hfil##\hfil$\crcr#2\crcr\sim\crcr}}}

\renewcommand{\thefootnote}{\fnsymbol{footnote}}

\begin{document} \begin{titlepage}
\setcounter{page}{1}
\thispagestyle{empty}
\rightline{\vbox{\halign{&#\hfil\cr
&SLAC-PUB-6672\cr
&September 1994\cr
&T/E\cr}}}
\vspace{0.8in}
\begin{center}

{\Large\bf
Searching for Anomalous Weak Couplings of Heavy Flavors at the SLC and LEP}
\footnote{Work supported by the Department of
Energy, contract DE-AC03-76SF00515.}
\medskip

\normalsize THOMAS. G. RIZZO
\\ \smallskip
{\it {Stanford Linear Accelerator Center\\Stanford University,
Stanford, CA 94309}}\\

\end{center}

\begin{abstract}

The existence of anomalous electric($\tilde \kappa$) and/or magnetic($\kappa$)
dipole moment couplings between the heavy flavor fermions ($c,b,\tau$) and
the $Z$ boson can cause significant shifts in the values of several
electroweak observables currently being probed at both the  SLC and LEP.
Using the good agreement between existing data and the predictions of the
Standard Model we obtain strict bounds on the possible strength of these new
interactions for all of the heavy flavors. The decay $Z\rightarrow b\bar b$,
however, provides some possible hint of new physics. The corresponding
anomalous couplings of $\tau$'s to photons is briefly examined.

\end{abstract}

\vskip0.45in
\begin{center}

Submitted to Physical Review {\bf D}.

\end{center}

%\noindent{(Talk given at the {\it Workshop on Photon Radiation from Quarks},
%Annecy, France, December 2-3, 1991.)}

\renewcommand{\thefootnote}{\arabic{footnote}} \end{titlepage}

%%%%%%%%%%%%%%%%%%%%%%%%%%%%%%%---- text

The Standard Model(SM) continues to provide an excellent description of almost
all aspects of existing experimental data{\cite {rosner}} especially in light
of the possible discovery of the top quark by the CDF Collaboration at the
Tevatron{\cite {cdf}} in the mass range anticipated by analyses of precision
electroweak data{\cite {lep}}. Even with its many successes we know the SM
cannot the be the whole story and the eventual discovery of new physics beyond
the SM has long been anticipated. Of course, we can not predict in what form or
exactly where this crack in the SM may first appear so all possible avenues to
its discovery must be explored. One possibility would be the production of a
new particle(\eg, a $Z'$) or set of particles(\eg, SUSY) not
contained within the SM framework at some high energy collider. A second
possibility might be the observation of a rare $K, D, \tau$ or $B$ decay either
forbidden by the SM or with a rate that is completely at variance with SM
expectations. A last, but still quite promising, possibility would be
a deviation from SM predictions in high precision measurements. The first
scenario obviously has the clear advantage in that the new physics is clean and
distinct whereas in the last two cases we perhaps learn little more that the
SM is incomplete. Of course, the new physics possibilities {\it are} limited
by the details of what is found experimentally. However, for many observables
it is likely that several different new physics scenarios could lead to the
same prediction and substantial analysis would be required in order to clarify
such a situation.

Among the possible ways new physics may manifest itself, one that has been
getting ever-increasing attention is the anomalous coupling of heavy flavor
fermions
to the conventional SM gauge bosons, \ie, $Z$, $W$, $\gamma$, and $g$. In the
case of a neutral, on-shell gauge boson, these anomalous couplings take the
form of either electric or magnetic dipole form factors; electric dipole
moments are inherently $CP$-violating. These two types of new couplings
represent the lowest dimensional
non-renormalizable operators which can be added to the usual SM Lagrangian
which signal new physics entering from some large mass scale. Since
the top is the heaviest fermion the presumption has been that its couplings
would be the most sensitive to the existence of this new high
mass scale physics.
{}From this expectation it follows that the possibility that the top may
possess
anomalous couplings has received the most attention in the
literature{\cite {htop,etop}}.
Of course one might extend this same argument
to {\it all} of the fermions of the third generation, $t$, $b$, $\tau$ and
perhaps as well to $c$. Since these particles have been around for quite
some time and
much data has been accumulated about their properties, it seems quite natural
to ask if these well-known heavy flavor fermions posses anomalous
couplings or, at the
very least, to ask what the limits are on such couplings from existing data.
$\tau$'s have in fact received some attention in this respect{\cite {tau}}
especially in regards to a possible $CP$-violation associated with an
electric dipole moment interaction with the $Z$. If such couplings {\it were}
ever to be found we would certainly need to investigate and understand
how they arose. Our approach to analyzing the effects of such hypothetical
couplings is purely phenomenological. We do not seek to address the possible
origin of these anomalous fermion couplings should they exist. We wish only
to examine how the new interactions would numerically modify electroweak
observables at SLD and LEP so that they can be discovered if they do indeed
exist.

In this paper we will make use of the latest data from both the LEP
Collaborations{\cite {lep}} and the SLD{\cite {sld}} to place simultaneous
constraints on the possible anomalous electric and magnetic dipole couplings
of $b$, $c$, and $\tau$ to the $Z$. (An early analysis along these lines
using only the data on the $Z$ partial widths and assuming only one of the two
possible anomalous couplings is nonzero at a time was presented
in {\cite {masso}}.)  The analysis presented below employs the
$Z\to b\bar b, ~c\bar c$ and $\tau \bar \tau$ partial widths together with the
corresponding forward-backward asymmetries, $A_{FB}$, from LEP as well as
the polarized forward-backward asymmetries, $A_{FB}^{pol}$, for $b$'s and
$c$'s from SLD. In the $\tau$ case, the integrated final state $\tau$
polarization, $P_{\tau}$, is also employed. In a additional separate
analysis for the
$\tau$, which tests $e-\mu-\tau$ universality, we also make use of the
corresponding partial width and forward-backward asymmetry data
for both $e$ and $\mu$ as well as the SLD measurement of the left-right
asymmetry, $A_{LR}$. Note that although we do not examine any $CP$-violating
asymmetry, we are still able to obtain a rather strong constraint on the
potential existence of electric dipole moments for these heavy fermions. The
limits we obtain on such couplings are either comparable to or superior than
those in the existing literature where $CP$-violating observables are
employed.

To begin, we need to define the generalized form of the interaction of the
heavy fermions with the $Z$ to set our normalization and other conventions.
If we define $\kappa$ and $\tilde\kappa$ the as real parts of the
magnetic and electric
dipole form factors evaluated at $q^2=M_Z^2$, our interaction Lagrangian can
be written as
\begin{equation}
{\cal L}={g\over {2c_w}}\bar f\left[\gamma_{\mu}
(v_f-a_f\gamma_5)+{i\over {2m_f}}
\sigma_{\mu\nu}q^{\nu}(\kappa_f^Z-i\tilde\kappa_f^Z\gamma_5)
\right]fZ^{\mu} \,,
\end{equation}
where $g$ is the standard weak coupling, $c_w=cos \theta_W$, $m_f$ is the
fermion mass, and $q$ is the $Z$'s four-momentum. At the $Z$ pole this
interaction leads to the following symbolic result for the $e^+e^-\to f\bar f$
unpolarized differential cross section at the tree level in the effective
Born approximation
\begin{eqnarray}
{1\over{\beta}}{d\sigma\over{dz}} &\sim& (v_e^2+a_e^2)\left[(v_f^2+a_f^2)
(1+\beta^2z^2)+(v_f^2-a_f^2)(1-\beta^2)\right] \nonumber \\
&+& 2\beta z (2v_ea_e)\left[2v_fa_f+2\kappa_f^Z a_f\right]\nonumber \\
&+& (v_e^2+a_e^2)\biggl[r_f\left[(\kappa_f^Z)^2+(\tilde \kappa_f^Z)^2\right]
(1-\beta^2z^2)+
(\kappa_f^Z)^2-(\tilde \kappa_f^Z)^2+4v_f\kappa_f^Z\biggr] \,,
\end{eqnarray}
where $z=cos \theta$, $r_f={M_Z^2\over {4m_f^2}}$, and
$\beta^2=1-{1\over {r_f}}$. It is important to notice that for all of the
heavy fermions under discussion $r_f$ is $O(10^2)$ or larger.
{}From this equation we can immediately write down a correspondingly symbolic
expression for the angular integrated
$Z\to f\bar f$ partial width, $\Gamma_f$, as
\begin{eqnarray}
{1\over {K\beta}}\Gamma_f &=& D=(v_f^2+a_f^2)(1+\beta^2/3)+(v_f^2-a_f^2)
(1-\beta^2)+r_f\left[(\kappa_f^Z)^2+(\tilde \kappa_f^Z)^2\right] \nonumber \\
&+& (\kappa_f^Z)^2-(\tilde \kappa_f^Z)^2+4v_f\kappa_f^Z \,,
\end{eqnarray}
where the overall normalization factor, $K$, is given by
\begin{equation}
K={N_cG_FM_Z^3\over {8\pi\sqrt{2}}} \,,
\end{equation}
with $N_c$ being the number of colors of the final state fermion, $G_F$, the
Fermi constant and $M_Z$, the $Z$ boson mass.
Similarly for the forward-backward asymmetry, we obtain
\begin{equation}
A_{FB}^f={3\over {4}}A_eA_f \,,
\end{equation}
where as usual
\begin{equation}
A_e={2v_ea_e\over {v_e^2+a_e^2}} \,,
\end{equation}
while for the heavy fermions with anomalous couplings, $A_f$ differs somewhat
from the usual expression in the SM. We find instead
\begin{equation}
A_f={4\over {3}}\beta (2v_fa_f+2\kappa_f^Za_f)/D \,,
\end{equation}
and $D$ is defined above. Of course, $A_f$ reverts to its usual form in the
limit that $\kappa_f^Z,\tilde\kappa_f^Z \to 0$, with $\beta \to 1$. The other
asymmetries are also easily found; $A_{LR}=A_e$ maintains its SM form while the
polarized forward-backward asymmetry can still be written in the SM form
\begin{equation}
A_{FB}^{pol}(f)={3\over {4}}A_f \,,
\end{equation}
with the value of $A_f$ now given as above. It is important to observe that
$A_{LR}$ is completely insensitive to the existence of any anomalous couplings
that might be possessed by the {\it final} state fermions.
To round out the usual list of
observables, we find that the expression for the angular averaged
polarization of the $\tau$ in $Z$ decay now can be expressed as
\begin{equation}
P_{\tau}=-{2v_\tau a_\tau(1+\beta^2/3)+2a_\tau\kappa_\tau^Z
\over {(v_\tau^2+a_\tau^2)(1+\beta^2/3)+
r_\tau\left[(\kappa_\tau^Z)^2+(\tilde \kappa_\tau^Z)^2\right](1-\beta^2/3)+
2v_\tau\kappa_\tau^Z}} \,,
\end{equation}
which reduces to the conventional SM result in the same limit as
described above. It is very important
to note that $P_{\tau} \neq -A_{\tau}$ even in the $\beta^2 \to 1$ limit
when either $\kappa_\tau^Z$ or $\tilde \kappa_\tau^Z$ is non-zero. (We will
assume in our analysis that no new physics enters the $\tau\nu_\tau W$ vertex
so that the measured values of $P_\tau$ can be interpreted as in the SM.)
Looking at
the expressions for the above observables we see that they are all
even functions
of $\tilde \kappa_f^Z$ while both even and odd terms in $\kappa_f^Z$ appear.
This is to be expected since $\tilde \kappa_f^Z$ is a coefficient of a
$CP$-violating interaction, \ie, the electric dipole moment operator. Since we
are only dealing with $CP$-even quantities the bounds we obtain on
$\tilde \kappa_f^Z$'s will be on their possible absolute values. Note
that in the
case of the $D$ parameter and in the denominator of the expression for
$P_\tau$, {\it both} quadratic $\tilde \kappa_f^Z$ and $\kappa_f^Z$ terms are
scaled by $r_f\geq 100$, as pointed out above, and which leads to a
significantly enhancement in the sensitivity of the various observables to both
these parameters. In particular, for the $b$ and $c$ cases this implies that
all these
observables are also, at least roughly, even functions of $\kappa_f^Z$.
This approximation will be badly violated in the case of the $\tau$
asymmetries since the SM term is suppressed by the small value of $v_\tau$
which thus allows the linear $\kappa_\tau^Z$ term in the numerators of both
asymmetry expressions to make a very considerable contribution.

How do the existence of non-zero values for $\kappa^Z$ and/or
$\tilde \kappa^Z$ numerically influence the $Z$-pole observables? Let us
first consider the case of $b$ and $c$ quarks where the three specific
quantities we will deal with are $R_{b,c}=\Gamma_{b,c}/\Gamma_{had}$,
$A_{FB}^{b,c}$, and $A_{FB}^{pol}(b,c)$. To define the predictions of the SM
we make use of the electroweak library in ZFITTER4.8{\cite {zfit}} which has
been augmented to include the recent $O(\alpha\alpha_s^2)$ results of
Avdeev, Fleischer, Mikhailov and Tarasov{\cite {fleisch}}. We fix the input
parameters as $M_Z=91.1888$ GeV, $M_H=300$ GeV, and $\alpha_s(M_Z)=0.125$
from Ref.3 and take three representative values of the top mass,
$m_t=165,~175$, and 185 GeV
to scan the range suggested by both electroweak fits{\cite {lep}} as well as
the CDF top search{\cite {cdf}}. (Our results are not particularly sensitive
to variations in the choices of $M_H$ or $\alpha_s(M_Z)$.) To include some of
the theoretical uncertainties into the analysis, we vary the default parameter
choices in ZFITTER and examine the spread in the predictions; we then use
the average of these values as the SM prediction and treat the standard
deviation from this average as a theory error which is included as an
additional uncertainty in the analysis below. (By varying these ZFITTER default
parameters we are allowing, \eg , for different treatments of the hadronic
vacuum polarization and different approaches for the resummation of
terms beyond those of leading order in $\alpha$.)

We begin with the charm case. Fig.1a shows the variation in the value of $R_c$
in comparison to that of the SM, for a fixed value of $m_t$, as either
$\tilde \kappa_c^Z$ {\it or} $\kappa_c^Z$ become non-zero. The three
predictions
corresponding to the three top mass choices lie underneath a single curve.
As expected, the $\tilde \kappa_c^Z$ result is symmetric under
$\tilde \kappa_c^Z \to -\tilde \kappa_c^Z$ while that for $\kappa_c^Z$ is
nearly so. $\tilde \kappa_c^Z$ or $\kappa_c^Z$ non-zero thus leads to an
{\it increase} in $R_c$ with values as small as 0.005 leading to a $4\%$
shift from the SM expectation. For $m_t$ fixed, determining the ratio of
either $A_{FB}^c$ or
$A_{FB}^{pol}(c)$ to their SM values amounts simply to calculating
$A_c/A_c^{SM}$ which we show in Fig.1b. This ratio exhibits a behaviour
similar to $R_c$ in its response to either $\tilde \kappa_c^Z$ or $\kappa_c^Z$
non-zero except that these now lead to a {\it decrease} in $A_c$. Values of
these
parameters of order 0.005 lead to shifts in $A_c$ of order $5\%$. Again, the
results for the three different $m_t$ choices lie beneath a single curve.
Combining the results of these two figures together we obtain Fig.1c which
also shows the data point obtained by combining the charm results from
LEP{\cite {lep}} on $R_c$ and $A_{FB}^c$ together with those from the
SLD{\cite {sld}} on $A_{FB}^{pol}(c)$, assuming $m_t=175$ GeV. The value
and error for
$A_c$ was obtained by combining the LEP and SLD results for the two
respective asymmetries. (The position of
the data point moves only slightly if the other $m_t$ values are assumed. As
we will see, this
makes little difference in the final results as the errors on the charm data
are still quite large.) The solid(dashed) curves in the upper portion of the
figure show the predictions when $\kappa_c^Z$($\tilde \kappa_c^Z$)
is non-zero with the diamonds representing steps in both these parameters in
units of 0.002. Since the central values of the width and asymmetry
measurements lie in the quadrant opposite to that predicted by non-zero values
of either $\tilde \kappa_c^Z$ or $\kappa_c^Z$, we anticipate that rather
strong bounds should be obtained. This is done by performing a $\chi^2$ fit
to the values of $R_c$, $A_{FB}^c$, and $A_{FB}^{pol}(c)$ for fixed $m_t$ and
allowing $\tilde \kappa_c^Z$ and/or $\kappa_c^Z$ to freely vary. Of course the
SM prediction itself is little more than about $1\sigma$ from the
central value of the data.
If we take $m_t=165, 175,$ or 185 GeV, and {\it assume } that
$\tilde \kappa_c^Z=0$, we find the following $95\%$ CL bounds on $\kappa_c^Z$:
($-5.8$ to 5.3)$\cdot 10^{-3}$, ($-5.9$ to 5.4)$\cdot 10^{-3}$,
($-6.0$ to 5.4)$\cdot 10^{-3}$. If on
the contrary we make the opposite assumption and take
$\kappa_c^Z=0$, we find that
$|\tilde \kappa_c^Z|<(5.6,5.7,5.8)\cdot 10^{-3}$ for the same $m_t$ choices,
respectively. If {\it both} $\tilde \kappa_c^Z$ and $\kappa_c^Z$ are permitted
to be non-zero simultaneously we obtain the $95\%$ CL region shown in Fig.1d.
Amongst other things this plot shows is that the absolute value of
the $c\bar cZ$ electric dipole
moment is $<5.7\cdot 10^{-17}$ e-cm in conventional units, independently of
whether a magnetic dipole moment {\it also} exists and independently of the
precise value of $m_t$. This result is a
significant improvement over that which was obtained previously{\cite {masso}}
with stronger assumptions. We note in passing that the positions of the three
$\chi^2$ minima correspond to $\tilde \kappa_c^Z=0$ with
$\kappa_c^Z=(-0.29^{+3.4}_{-3.1}, -0.35^{+3.3}_{-3.1}, -0.42^{+3.3}_{-3.2})
\cdot 10^{-3}$, all of which lie quite close to the origin in Fig.1d.

Let us now turn to the case of $b$'s where we will follow a similar
procedure. Due to the larger $b$-quark mass we anticipate somewhat poorer
limits than what was obtained in the charm case. However, a new wrinkle emerges
in the $b$ case in that the existing data {\it prefer} non-zero values for
the anomalous couplings.
The $Z\to b\bar b$ situation is, of course, very interesting since it has been
known for some time that $R_b$ lies{\cite {lep}} about $2\sigma$ above its
SM predicted value. For $m_t=175$ GeV, the LEP value of $A_{FB}^b$ is about
$1\sigma$ low and the value of $A_{FB}^{pol}(b)$ from SLD{\cite {sld}} also
lies a bit below the SM prediction but with larger errors.
Figs.2a and 2b show how both $R_b$ and $A_b$ vary with either
$\tilde \kappa_b^Z$ or $\kappa_b^Z$ non-zero; again, in both cases the results
for the three different values of the top mass lie underneath a single curve.
Both observables are found to have comparable sensitivity to the existence
of anomalous couplings.
Note that non-zero values of either $\tilde \kappa_b^Z$ or $\kappa_b^Z$ will
push the SM predictions closer to the data, \ie, they lower $A_b$ while
increasing
$R_b$. This is perhaps seen more clearly in Fig.2c which not only shows the
model predictions but also the data points for $m_t$=165, 175 or 185 GeV.
Anomalous couplings will certainly lead to a better fit than does the SM. Let
us first assume that $\tilde \kappa_b^Z=0$ and find the allowed ranges for
$\kappa_b^Z$. For $m_t=165$ GeV the $95\%$ CL allowed range is determined to be
($-1.11$ to $-0.03$)$\cdot 10^{-2}$; the SM point lies $2.1\sigma$ away from
the
minimum, just outside the $95\%$ CL allowed range. For $m_t=175$ GeV, a
secondary $\chi^2$ minimum develops so that the allowed ranges for
$\kappa_b^Z$ are ($-1.18$ to $-0.14$)$\cdot 10^{-2}$ or
(2.28 to 2.95)$\cdot 10^{-2}$ and the SM lies more than $2.3\sigma$ away from
the $\chi^2$ minimum best fit.
Similarly, for $m_t=185$ GeV, the allowed $\kappa_b^Z$ ranges
are ($-1.23$ to $-0.24$)$\cdot 10^{-2}$ and (2.21 to 3.15)$\cdot 10^{-2}$ with
the
SM now being about $2.6\sigma$ away from the corresponding best fit.
A similar situation
occurs in the reverse case where we assume that $\kappa_b^Z=0$ and we look for
the $95\%$ CL bounds on $|\tilde \kappa_b^Z|$. For $m_t=165, 175, 185$ GeV,
the allowed ranges are (0.15 to 1.89)$\cdot 10^{-2}$,
(0.66 to 1.99)$\cdot 10^{-2}$, and (0.85 to 2.07)$\cdot 10^{-2}$. The SM point
lies $2.0\sigma$, $2.4\sigma$, and $2.8\sigma$ away from the $\chi^2$ minima
in these three cases, respectively.

If we allow both $\tilde \kappa_b^Z$
{\it and} $\kappa_b^Z$ to be present simultaneously we arrive at the plot
shown in Fig.2d. Note that for $m_t>175$ GeV the SM lies outside of the
two parameter $95\%$ CL region. Clearly we cannot yet make any claim for the
existence of new physics but it is clear from this analysis that the
observables associated with the $Z\to b\bar b$ mode should be closely
monitored.
If we {\it ignore} the hole near the origin, the limits obtained above on
$\tilde \kappa_b^Z$ can be re-written in more conventional units at
$\tilde \kappa_b^Z \leq 6.0\cdot 10^{-16}$ e-cm.
For completeness, we note the approximate positions of the $\chi^2$ minima
for $m_t=165$, 175, and 185 GeV: ($\kappa_b^Z$, $|\tilde \kappa_b^Z|)
=(-6.61\cdot 10^{-3},0), ~(3.64\cdot 10^{-3}, 1.67\cdot 10^{-2})$, and
$(8.56\cdot 10^{-3}, 1.86\cdot 10^{-2})$, respectively.

We now turn to the case of $\tau$'s. There are two possible approaches: ($i$)
one can simply follow the same approach as employed above for $c$ and $b$
with the substitution of $P_\tau$ for $A_{FB}^{pol}(b,c)$ as long as we
remember that $P_\tau \neq -A_\tau$. We call this the `standard' approach.
In principle we might also include the
additional constraint arising from the full angular dependent $\tau$
polarization(as conventionally represented by the `$A_e$' term). However,
this extra information is
obtained under the assumption of a specific form of the angular dependence of
the $\tau$ polarization which is somewhat modified when anomalous couplings are
present. To avoid this complexity we will not include the angular dependent
information in the present analysis. The additional constraints obtainable
by its inclusion are not, however, expected to be significant since the
{\it odd} term in the angular distribution is only weakly dependent on the
existence
of anomalous couplings due to its proportionality to $v_\tau$. A more
general approach to handling the full angular
dependence of the $\tau$ polarization is now underway.

A second
possibility is to redefine what we mean by the SM prediction for the various
observables. In this approach, we assume $e-\mu$ universality and use the other
leptonic data from LEP and SLD to {\it define} the SM prediction. We call this
the `universality' approach.
As an example,
we now define the SM prediction for the $Z\to \tau \bar \tau$ partial width
to be the error weighted average of the the $Z\to e\bar e$ and
$Z\to \mu \bar \mu$ partial widths corrected for the $\tau \bar \tau$ phase
space. The resulting bounds we obtain in this approach are very insensitive
to the values we assume for the top mass, $m_t$.

Taking the `standard' approach, we plot in Figs.3a-c the variation in the SM
prediction for the
$Z\to \tau \bar \tau$ partial width($\Gamma_\tau$), the $\tau$ forward-backward
asymmetry and the $\tau$ polarization when either $\tilde \kappa_\tau^Z$ or
$\kappa_\tau^Z$ is non-zero for the usual three choices of $m_t$. In Fig.3a,
as usual, we see that there is no observable sensitivity to the choice of
$m_t$ when the ratio to the SM prediction is taken for either $\kappa_\tau^Z$
or $\tilde \kappa_\tau^Z$ non-zero. Since $v_\tau$ is very small and $r_\tau$
is so large the $\kappa_\tau^Z$ non-zero curve is almost perfectly symmetric
about the origin. The value of $\Gamma_\tau$ varies only a few per cent as
the anomalous couplings range over $\pm 0.005$.
In both Figs.3b and 3c we see something different than in
all of the other results so far obtained, \ie, the $\tilde \kappa_\tau^Z$
and $\kappa_\tau^Z$ non-zero scenarios behave very differently. This is due
to ($i$) the presence of $\kappa_\tau^Z$ (but not $\tilde \kappa_\tau^Z$ !)
appearing linearly in the numerators for the expressions of both $A_\tau$ and
$P_\tau$ and ($ii$) the fact that $v_\tau$ is very small. In these two figures
we also note for the first time a barely observable separation between the
$m_t=165$, 175 and 185 GeV model predictions. We also observe that the
asymmetries
are quite sensitive to anomalous couplings with variations as large as $10\%$
away from SM expectations.

In Fig.3d we compare the shifts in the $Z\to \tau\bar \tau$ partial width
and $P_\tau$ for non-zero values of $\tilde \kappa_\tau^Z$ or $\kappa_\tau^Z$
since these are at present the most accurately determined quantities; steps
of 0.001 are indicted by the diamonds. (Note that close to the SM point there
is no separation in the $m_t=165$, 175, and 185 GeV predictions.) The results
of the LEP measurements normalized to the SM expectations with $m_t=175$ GeV is
also shown. If we perform a $\chi^2$ fit to the data, limits on both
$\tilde \kappa_\tau^Z$ and $\kappa_\tau^Z$ are obtained in the usual manner.
For $\tilde \kappa_\tau^Z=0$, we obtain the following $95\%$ CL ranges for
$\kappa_\tau^Z$: ($-3.04$ to 0.88)$\cdot 10^{-3}$,
($-2.90$ to 2.09)$\cdot 10^{-3}$, and ($-2.75$ to 2.32)$\cdot 10^{-3}$ for
$m_t=165$, 175 and 185 GeV respectively. In the reverse case, the
corresponding upper bounds on $|\tilde \kappa_\tau^Z|$ are found to be
(2.9, 3.0, 3.1)$\cdot 10^{-3}$. If both $\tilde \kappa_\tau^Z$ and
$\kappa_\tau^Z$ are non-zero we obtain the allowed regions shown in Fig.3e.
The $\chi^2$ minima all occur at $\tilde \kappa_\tau^Z=0$ with
$\kappa_\tau^Z=(-1.95^{+0.81}_{-0.60}, -1.74^{+0.95}_{-0.65},
-1.50^{+1.20}_{-0.72})\cdot 10^{-3}$, respectively.
Note that for any value of $m_t$, we obtain a $95\%$ CL limit on $\tau$
electric dipole moment of less than $2.1\cdot 10^{-17}$ e-cm which is quite
comparable to that obtained by the OPAL Collaboration{\cite {thesis}} through
the use of $CP$-violating observables.

Turning now to the universality approach, we repeat the analysis above using
the LEP and SLD $e$ and $\mu$ data to define the SM predictions. Of course
the results in Figs.3a-d are unmodified with only the position of the data
point changing in Fig.3d to reflect the change in the SM prediction. The
result of this analysis leads to the additional curve in Fig.3e where we see
that results comparable to but a bit weaker than the conventional
analysis are obtained. The $\chi^2$
minimum now occurs at ($\kappa_\tau^Z$, $|\tilde\kappa_\tau^Z|$)=
($0.29^{+1.52}_{-1.66}$, $1.87^{+1.79}_{-0.78}$)$\cdot 10^{-3}$. For
$\tilde \kappa_\tau^Z=0$, we obtain the
$95\%$ CL range for $\kappa_\tau^Z$: ($-2.96$ to 3.27)$\cdot 10^{-3}$ and, for
$\kappa_\tau^Z=0$, we obtain the bound
$|\tilde \kappa_\tau^Z| \leq 3.20\cdot 10^{-3}$.

Interestingly, a procedure similar to that above which employs lepton
universality can be used to obtain
reasonably strong limits on the corresponding anomalous
$\tau\bar \tau \gamma$ couplings{\cite {tc}}. If we compare the three cross
sections for $e^+e^- \to e^+e^-,\mu^+\mu^-$, and $\tau\bar \tau$ at TRISTAN
energies{\cite {tristan}}(and properly subtract
out the $t-$channel pole in the $e^+e^-$ case), limits on universality
violation can be used to place constraints on $\tilde \kappa_\tau^\gamma$ and
$\kappa_\tau^\gamma$ {\it provided} we assume that the anomalous $Z$
couplings can be neglected. Fig.4 shows a comparison of the $R$ ratio expected
in the $\tau$ case with anomalous couplings to that of the SM assuming
universality but with finite $\tau$ mass corrections for TRISTAN energies.
At an average center of mass energy
$\sqrt {s} \simeq 57.8$ GeV, these ratios are very well
determined{\cite {tristan}} and we find from the $95\%$ CL upper limit on
the ratio
$R_\tau/R_{e\mu}<1.10$ that $|\tilde \kappa_\tau^\gamma| \leq 2.8\cdot 10^{-2}$
when $\kappa_\tau^\gamma=0$ and correspondingly, when
$\tilde \kappa_\tau^\gamma=0$, $\kappa_\tau^\gamma$ lies in the $95\%$ CL
interval ($-1.9$ to 4.2)$\cdot 10^{-2}$. (These constraints may be improved
somewhat
by using additional data such as $A_{FB}^\tau$.)
Although these bounds are inferior to those
obtained on the corresponding $Z$ couplings (due to lower statistics and
reduced sensitivity), they are substantially better than those found
in the existing literature which were arrived at by other methods. For
example, Grifols and Mendez{\cite {grif}}
examined the radiative decay $Z\to \tau\bar \tau \gamma$ and obtained an upper
bound on $|\kappa_\tau^\gamma|$ of 0.11 by looking for excess events.

In this paper we have undertaken a systematic search for the effects of
anomalous electric or magnetic dipole moment type couplings between the heavy
flavor fermions and the $Z$ based on precision data from the SLC and LEP. For
the $b$ and $c$ quarks $R_{b,c}$, $A_{FB}^{b,c}$, and $A_{FB}^{pol}(b,c)$
were used {\it simultaneously} to obtain our results. In the $\tau$ case, our
first approach followed that for the quark case but replaced $A_{FB}^{pol}$
with $P_\tau$ while our second approach
employed lepton universality. The results of this analysis can be summarized as
follows:

($i$) The constraints we obtained on the anomalous couplings of the $\tau$ and
charm were found to be reasonably insensitive to the details of the SM
radiative corrections which were expressed via variations in $m_t$. This was
quite forcefully demonstrated in the $\tau$ case where the universality limit
was used as the reference SM. All of the observables played a role in
obtaining the allowed ranges. The individual numerical results are summarized
for comparison in Tables 1 and 2. It is important to note that the constraints
obtained in the charm case are inferior to those obtained for $\tau$'s even
though they have comparable masses. Of course, the data in the case for
$\tau$'s is more precise which is the major source of the difference. In the
$b$ case, the larger fermion mass reduces sensitivity while the data itself
shows some preference for the existence of anomalous couplings.
The limits we obtain are generally
stronger than those derived
previously and neither $\tau$'s nor charm showed any indication of anomalous
couplings. The inclusion of the angular-dependent $\tau$ polarization data
from LEP is not expected to make any significant effect on these results but
is a subject of further study.

($ii$) The situation in the $b$ case is quite different than either charm or
$\tau$ in that it shows a much greater sensitivity to variations in $m_t$
and that non-zero values of the anomalous couplings are somewhat more favored
by the fits. For example, with $m_t=175$ GeV, the SM lies just outside the
$95\%$ CL region in the two-parameter fits and a bit further outside this CL
range for the two, one-parameter fits(see Tables 1 and 2).
The reason for this is immediately clear from Fig.2c, \ie,
the presence of the anomalous couplings induces a larger value for $R_b$ while
simultaneously decreasing $A_b$, which is just the direction taken by
the present data. Although we
can make no claim for new physics at the current level of statistics it is
clear that all observables related to the decay $Z\to b\bar b$ should be
watched carefully and scrutinized.

($iii$) The universality approach for $\tau$'s was extended to the $\gamma$
case using TRISTAN data under the assumption that the contribution of the
corresponding anomalous $Z$ couplings were suppressed. The limits so obtained
are a significant improvement over those already existing in the literature.

The possible existence anomalous couplings of the heavy fermions to the $Z$
may provide a clue to new physics beyond the Standard Model.

\vskip.25in
\centerline{ACKNOWLEDGEMENTS}

The author like to thank A. Kagan, D. Atwood, J.L. Hewett, T. Barklow,
S. Wagner, T. Junk, P. Rowson, M. Hildreth and P. Burrows for discussions
related to this work. Special thanks go to S. Wagner for a careful reading
of the manuscript.
He would also like to thank the members of the Argonne National Laboratory
High Energy Theory Group for use of their computing facilities.

\newpage

%
%%%%%%%%%%%%%%%%%%--- References
%%%%%%%%%%%%%%%%%%%%%%%%%%%%%%%%%%%%%%%%%%%%%%%%%%%%%%%
\def\MPL #1 #2 #3 {Mod.~Phys.~Lett.~{\bf#1},\ #2 (#3)}
\def\NPB #1 #2 #3 {Nucl.~Phys.~{\bf#1},\ #2 (#3)}
\def\PLB #1 #2 #3 {Phys.~Lett.~{\bf#1},\ #2 (#3)}
\def\PR #1 #2 #3 {Phys.~Rep.~{\bf#1},\ #2 (#3)}
\def\PRD #1 #2 #3 {Phys.~Rev.~{\bf#1},\ #2 (#3)}
\def\PRL #1 #2 #3 {Phys.~Rev.~Lett.~{\bf#1},\ #2 (#3)}
\def\RMP #1 #2 #3 {Rev.~Mod.~Phys.~{\bf#1},\ #2 (#3)}
\def\ZP #1 #2 #3 {Z.~Phys.~{\bf#1},\ #2 (#3)}
\def\IJMP #1 #2 #3 {Int.~J.~Mod.~Phys.~{\bf#1},\ #2 (#3)}

\newpage

\vskip2.0in

\vglue 1.0in
\begin{table}
\centering
\begin{tabular}{|l|c|c|c|} \hline\hline
$m_t$ & $c$ & $b$ & $\tau$ \\ \hline\hline
165 & $-5.8$ to 5.3 & $-11.1$ to $-0.3$ &$-3.04$ to 0.88 \\
175 & $-5.9$ to 5.4 & $-11.8$ to $-1.4$ &$-2.90$ to 2.09 \\
& & ~22.8 to 29.5 &\\
185 & $-6.0$ to 5.4 & $-12.3$ to $-2.4$ &$-2.75$ to 2.32 \\
& & ~22.1 to 31.5 &\\
 U  & -- & -- &$-2.96$ to 3.27 \\ \hline\hline
\end{tabular}
\caption{Individual $95\%$ CL allowed ranges for $\kappa^Z$ in units of
$10^{-3}$. `U'
corresponds to the universality approach for $\tau$'s described in the text.}

\end{table}
\vglue 1.0in
\begin{table}
\centering
\begin{tabular}{|l|c|c|c|} \hline\hline
$m_t$ & $c$ & $b$ & $\tau$ \\ \hline\hline
165 & $<5.6$ & 1.5 to 18.9 &$<2.9$ \\
175 & $<5.7$ & 6.6 to 19.9 &$<3.0$ \\
185 & $<5.8$ & 8.5 to 20.7 &$<3.1$ \\
 U  & -- & -- &$<3.2$ \\ \hline\hline
\end{tabular}
\caption{Individual $95\%$ CL allowed ranges for $|\tilde \kappa^Z|$ in
units of
$10^{-3}$. `U' corresponds to the universality approach for $\tau$'s as
described in the text.}
\end{table}
\vspace{2.0in}
\newpage

%%%%%%%%%%%%%%%%%%%%%%%--- figures
%
{\bf Figure Captions}
\begin{itemize}

\item[Figure 1.]{(a)$R_c$ and (b)$A_c$ variations due to non-zero values for
either $\tilde \kappa_c^Z$(dashes) or $\kappa_c^Z$(solid). The predictions for
$m_t=165$, 175, and 185 GeV lie underneath a single curve.
(c)$R_c$ vs. $A_c$ for non-zero values of $\tilde \kappa_c^Z$(dashed) or
$\kappa_c^Z$(solid) in comparison to the LEP and SLD data which is plotted
assuming $m_t=175$ GeV in the calculations of the SM predictions. As noted in
the text, the position and error associated with $A_c$ comes from combining
the SLD results for $A_{FB}^{pol}$ as well as the LEP determinations of
$A_{FB}^c$. The diamonds
correspond to incremental changes in
either $\tilde \kappa_c^Z$ or $\kappa_c^Z$ away from zero in steps of 0.002.
The upper(lower) solid curve is for $\kappa_c^Z$ positive(negative).
(d)$95\%$ CL allowed region in the $\tilde \kappa_c^Z$-$\kappa_c^Z$
plane resulting
from a fit to LEP data on $R_c$ and $A_{FB}^c$ and SLD data on
$A_{FB}^{pol}(c)$ with $m_t=$165(dashed), 175(solid), or 185(dotted) GeV. The
allowed region lies between the lower axis and the particular curve.}
\item[Figure 2.]{Same as Figs.1a-d but with $c \to b$. In (c) the data is
now displayed assuming SM predictions with $m_t=165$(dashes), 175(solid) or
185(dotted) GeV. The diamonds are now increments of 0.02 in either
$\tilde \kappa_b^Z$ or $\kappa_b^Z$ with the upper solid curve corresponding
to {\it negative} values of $\kappa_b^Z$. The allowed region in (d) lies
between pairs of curves of similar type.}
\item[Figure 3.]{Variation in the SM predictions for (a) the $\tau$ partial
width of the $Z$, (b)the $\tau$ forward-backward asymmetry and (c) final
state $\tau$ polarization for non-zero values of either
$\tilde \kappa_\tau^Z$ or $\kappa_\tau^Z$. In (a) the solid(dashed) curve
corresponds to variations in $\kappa_\tau^Z$($\tilde \kappa_\tau^Z$) for the
usual three choices of $m_t$. In (b) and (c) the $\tilde \kappa_\tau^Z$
non-zero case
corresponds to the horizontal dashed curve for all three values of $m_t$
whereas the steeper dashed(solid,dotted) curves correspond to non-zero
$\kappa_\tau^Z$ with $m_t=165(175,185)$ GeV. (d) $\Gamma_\tau$ vs $P_\tau$
with $\tilde \kappa_\tau^Z$($\kappa_\tau^Z$) varying in steps of 0.001 along
the dashed(dotted) curve with positive(negative) values of $\kappa_\tau^Z$
being to the left(right) of the SM point. The combined LEP result assuming
$m_t=175$ GeV in the SM calculation is shown as the data point. (e) Same as
Fig.1d but now for the case of $\tau$'s. The outer square dotted curve is the
resulting bound obtained from the $e-\mu-\tau$ universality analysis. The
allowed region lies below the curves.}
\item[Figure 4.]{The ratio of the $e^+e^-\to \tau\bar \tau$ cross section to
that for $e^+e^-\to e^+e^-,\mu^+\mu^-$ at TRISTAN
energies($\sqrt {s}\simeq 57.8$ GeV)  as a function of
$\tilde \kappa_\tau^\gamma$(dots) or $\kappa_\tau^\gamma$(solid). The
horizontal dashed line represents the $95\%$ CL upper limit on the cross
section ratio.}
\end{itemize}


\begin{thebibliography}{99}
\bibitem{rosner}
J.L.\ Rosner, summary talk presented at the {\it Eighth Meeting of the
Division of Particles
and Fields}, Albuquerque, NM, August 2-6, 1994. See also, J.L.\ Rosner,
EFI report EFI 94-38,
1994.
\bibitem{cdf}
F.\ Abe \etal, CDF Collaboration, \PRD D50 2966 1994 .
\bibitem{lep}
D.\ Schaile, talk presented at, and LEP reports LEPEWWG/94-02, LEPTAU/94-02,
LEPLINE/94-01, and LEPHF/94-03,
submitted to the {\it $27^{th}$ International Conference on High Energy
Physics}, Glasgow, Scotland, July 20-27, 1994.
\bibitem{htop}
G.\ Kane, G.A.\ Ladinsky and C.P.\ Yuan, \PRD D45 124 1992 ;
C.P.\ Yuan, \PRD D45 782 1992 ;
D.\ Atwood, A.\ Aeppli and A.\ Soni, \PRL 69 2754 1992 ;
D.\ Atwood, A.\ Kagan and T.G.\ Rizzo, SLAC report SLAC-PUB-6580, 1994.
\bibitem{etop}
M.\ Peskin, talk presented at the {\it Second International Workshop on Physics
and Experiments at Linear $e^+e^-$ Collider},  Waikoloa, HI, April 1993;
M.\ Peskin and C.R.\ Schmidt, talk presented at the {\it First Workshop on
Linear Colliders}, Saariselk\" a, Finland, September 1991; P.\ Zerwas, \ibid;
W.\ Bernreuther \etal, in {\it Proceedings of
the Workshop on \epem\ Collisions at 500 GeV, The Physics Potential},
(DESY, Hamburg) ed.\ by P.\ Igo-Kemenes and J.H.\ K\" uhn, 1992; A.\ Djouadi,
ENSLAPP-A-365-92 (1992);
M.\ Frigeni and R.\ Rattazzi, \PLB B269 412 1991 ;
R.D.\ Peccei, S.\ Persis and X.\ Zhang, \NPB B349 305 1991 ;
D.O.\ Carlson, E.\ Malkawi and C.-P.\ Yuan, Michigan State report
MSUHEP-94/05, 1994;
T.G.\ Rizzo, SLAC report SLAC-PUB-6512, 1994. Indirect
limits on $\kappa_t^g$
and $\tilde \kappa_t^g$ from the decay $b \to s \gamma$ were considered by
J.L.\ Hewett and T.G.\ Rizzo, \PRD D49  319 1994 .
\bibitem{tau}
A.\ Grifols and A.\ Mendez, \PLB B255 611 1991 ~and
erratum \PLB B259 512 1991 ;
B.\ Ananthanarayan and S.D.\ Rindani \PRL 73 1215 1994 ;
F.\ del Aguila and M.\ Sher, \PLB B252 116 1990 ;
R.\ Escribano and E.\ Masso, \PLB B301 419 1993 ;
W.\ Bernreuther, O.\ Nachtmann and P.\ Overmann, \PRD D48 78 1993 ;
G.\ Couture, \PLB B305 306 1993  ~and \PLB B272 404 1991 ;
G.\ Domokos \etal, \PRD D32 247 1985 ;
J.\ Reid, M.\ Samuel, K.A.\ Milton and T.G.\ Rizzo, \PRD D30 245 1984 .
See also, P.D.\ Acton \etal, OPAL Collaboration, \PLB B281 305 1992 ;
D.\ Buskulic \etal, ALEPH Collaboration, \PLB B297 459 1992 .
\bibitem{sld}
K.\ Abe \etal, SLD Collaboration, \PRL 73 25 1994  ~and SLAC report
SLAC-PUB-6607, 1994.
See also the presentations by M.E.\ King, D.\ Williams and G.\ Baranko, SLD
Collaboration, given at the {\it Eighth Meeting of the Division of Particles
and Fields}, Albuquerque, NM, August 2-6, 1994 and by S.\ Wagner at the
{\it $27^{th}$ International Conference on High Energy
Physics}, Glasgow, Scotland, July 20-27, 1994.
\bibitem{masso}
R.\ Escribano and E.\ Masso, Barcelona report UAB-FT-317, 1993.
\bibitem{zfit}
The ZFITTER package: D.\ Bardin \etal, \ZP C44 493 1989 ;
\NPB B351 1 1991  ; \PLB B255 290 1991  ; CERN report CERN-TH-6443/92, 1992.
The author would like to thank S. Riemann for providing him with an updated
version of this code.
\bibitem{fleisch}
L.\ Avdeev, J.\ Fleischer, S.\ Mikhailov and O.\ Tarasov, Univ. of Bielefeld
report BI-TP-93/60, 1994.
\bibitem{thesis}
A.\ H\" ocker, OPAL Collaboration, Universit\" at Bonn thesis, BONN-IB-94-17,
May 1994.
\bibitem{tc}
T.G.\ Rizzo, talk given at the {\it Workshop on the Tau-Charm Factory in the
Era of B-Factories and CESR}, SLAC, August 15-16, 1994.
\bibitem{tristan}
C.\ Velissaris, \etal, AMY Collaboration, \PLB B331 227 1994 ;
B.\ Howell, \etal, TOPAZ Collaboration, \PLB B291 206 1992 ;
K.\ Abe, \etal, VENUS Collaboration, \ZP C48 13 1990 .
\bibitem{grif}
See the analysis by A.\ Grifols and A.\ Mendez in Ref.6.

\end{thebibliography}
\end{document}